\documentclass[intlimits,twoside,a4paper]{article}

\usepackage{amsmath,amssymb}
\usepackage{graphicx}
\usepackage{wrapfig}

\usepackage[T2A]{fontenc}
\usepackage[cp1251]{inputenc}
\usepackage{cmpj2}

\issue{2012}{15}{3}{33704}
\doinumber{10.5488/CMP.15.33704}

\title[Exciton spectrum in multi-shell hexagonal semiconductor nanotube]%
{Exciton spectrum in multi-shell hexagonal semiconductor nanotube}
\author[O.M.~Makhanets \textsl{et al.} ]{O.M.~Makhanets\footnote{E-mail: ktf@chnu.edu.ua}\, , V.I.~Gutsul,  N.R.~Tsiupak, O.M.~Voitsekhivska}
\address{Chernivtsi National University, 2 Kotsyubinsky Str.,
58012 Chernivtsi, Ukraine}
\date{Received March 16, 2012, in final form May 29, 2012}

\authorcopyright{O.M.~Makhanets, V.I.~Gutsul,  N.R.~Tsiupak, O.M.~Voitsekhivska, 2012}

\begin{document}

\maketitle

\begin{abstract}
The theory of exciton spectrum in multi-shell hexagonal
semiconductor nanotube is developed within the effective masses
and rectangular potentials approximations using the method of
effective potential. It is shown that the exciton binding energy
for all states non-monotonously depends on the inner wire diameter,
approaching several minimal and maximal magnitudes. The obtained
theoretical results explain well the experimental positions of
luminescence peaks for GaAs/Al$_{0.4}$Ga$_{0.6}$As nanotubes.
\pacs 73.21.Hb, 78.67.Ch, 78.67.Lt
\end{abstract}

\section{Introduction}

\looseness=-1The semiconductor quantum wires are theoretically and
experimentally studied during more than twenty years. The improved
methods of their growth (molecular beam, gas phase and metal
organic epitaxy) gave an opportunity to produce arrays of quantum
nanowires with a radial  heterostructure~\cite{1,2}.

On the one hand, the heterostructure perpendicular to the quantum
wire axis can localize the charge carriers inside the inner
wire, thus decreasing the surface scattering~\cite{3}. On the other hand, this
allows a guided change of spectral parameters of quasi-particles (electrons, excitons,
phonons) depending on nanostructure geometric
parameters. The unique properties of quasi-particles make it possible to
utilize such systems as the basic elements of tunnel
nanodiodes, nanotransistors with high mobility of electrons,
effective light emitting devices, photo electric transformers,
nanosensors used for the diagnostics of biological and chemical
compositions~\cite{4}.

One of the variety of quantum wires with radial heterostructure is
a semiconductor nanotube intensively investigated recently. The
single (with one quantum well for quasi-particles)~\cite{5,6} and complex multi-shell (with several wells)~\cite{7,8} hexagonal
nanotubes are already produced experimentally using different
semiconductor materials.

The investigation of exciton binding energy for these structures
encounters serious mathematical problems connected with the necessary
correlation of spherical symmetry of Coulomb potential describing
the electron-hole interaction and non-spherical symmetry of a system
itself. Therefore, the exciton spectrum is often studied within
different and rather simple variational methods~\cite{9,10}, which are capable of quite well describing
only the exciton ground state. The method of effective potentials turns out to be more
informative. The theory of exciton spectrum in single cylindrical semiconductor
quantum wires has already been developed~\cite{11,12} using this method.

In this paper we propose one of the theoretical approaches to
the solution of the problem of exciton spectrum in multi-shell hexagonal
semiconductor nanotube. The theory is developed within the model
of effective masses and rectangular potentials using the method of
effective potentials. We study the parameters of exciton spectra
depending on the geometrical parameters of a nanostructure and
compare the numerical results obtained for the exciton energies
with the positions of luminescence peaks observed experimentally~\cite{8}.

\section{Theory of exciton spectra in multi-shell hexagonal semiconductor na\-no\-tube}

\begin{wrapfigure}{o}{0.5\textwidth}
\centerline{
\includegraphics[width=0.49\textwidth]{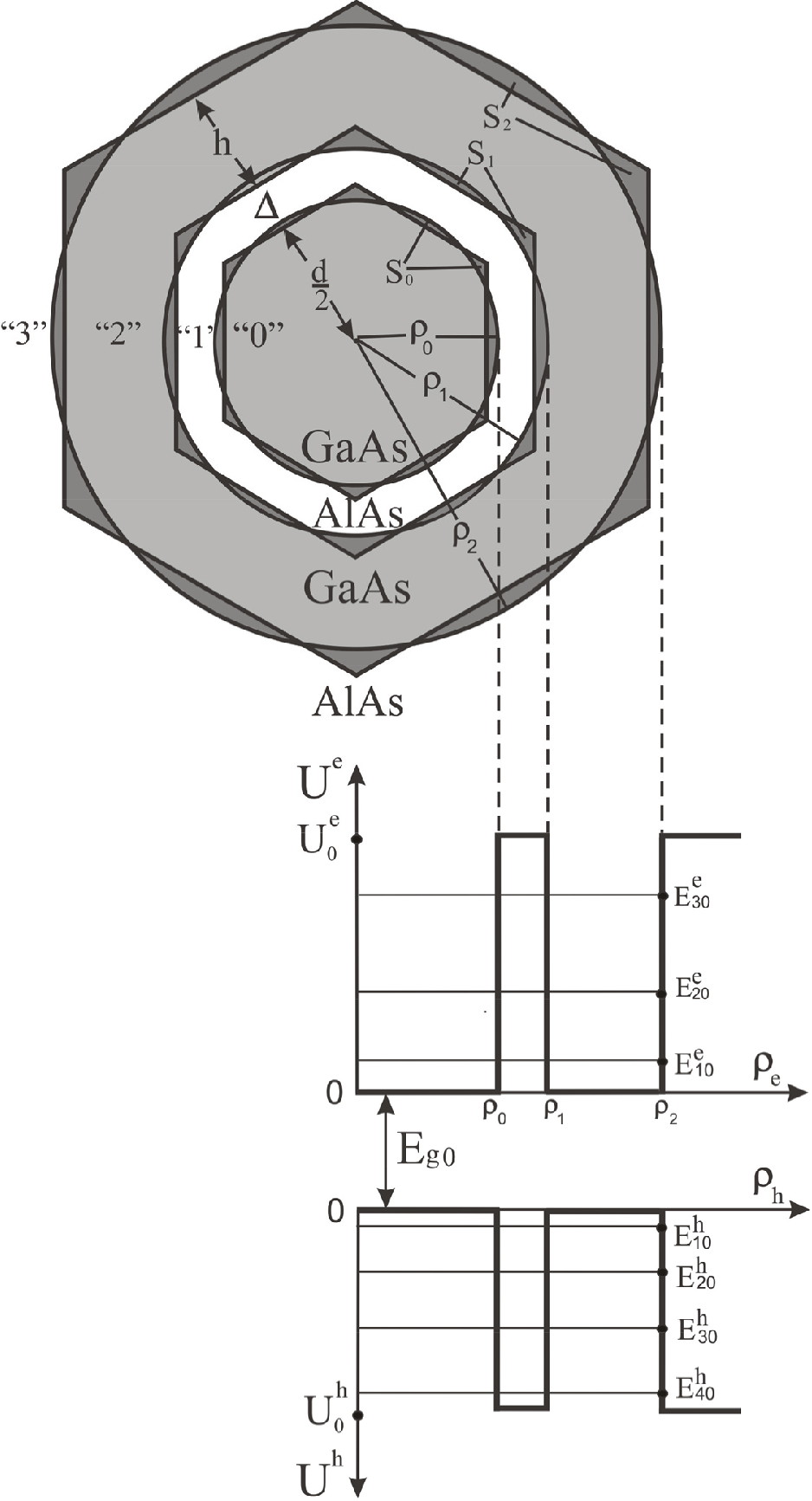}
}
\caption{Transversal cross-section of nanostructure and its energy
scheme.}\label{fig1}
\vspace{-1.5cm}
\end{wrapfigure}

The experimentally grown nanostructure~\cite{8}~-- the multi-shell
hexagonal nanotube is theoretically studied. It consists of hexagonal
semiconductor quantum wire (``0''), thin barrier-shell (``1'') and
nanotube (``2'') embedded into the outer medium (``3''). The
transversal cross-section of nanostructure is shown in figure~\ref{fig1}.

%
Within the assumption that the lattice and dielectric constants of
nanostructure compositions do not differ much, we use the
effective masses ($m^\mathrm{e,h}$) and rectangular potentials
($U^\mathrm{e,h}$) models to calculate the electron, hole and
exciton spectra.

We solve the Schr\"odinger equation for the exciton
\begin{equation}
\label{eq1}
\hat{H} _\mathrm{ex} (\vec {r}_\mathrm{e} ,\,\vec {r}_\mathrm{h} )\,\Psi
_\mathrm{ex} (\vec {r}_\mathrm{e} ,\,\vec {r}_\mathrm{h} ) = E_\mathrm{ex} \,\Psi _\mathrm{ex} (\vec
{r}_\mathrm{e} ,\,\vec {r}_\mathrm{h} )
\end{equation}
with the Hamiltonian
\begin{equation}
\label{eq2}
 \hat {H} _\mathrm{ex} (\vec {r}_\mathrm{e} ,\,\vec {r}_\mathrm{h} ) =
E_{g\,0} + \hat {H} ^\mathrm{e} (\vec {r}_\mathrm{e} )\, + \hat {H} ^\mathrm{h} (\,\vec
{r}_\mathrm{h} ) + V(\vert \vec {r}_\mathrm{e} - \vec {r}_\mathrm{h} \vert ){\rm ,}
\end{equation}
where $E_{g0}$ is the band gap of the nanotube,
\begin{align}\label{eq3}
&\hat {H} ^{(p)}= - {\frac{{\hbar ^{2}}}{{2}}}\left[ \vec
{\nabla} _{\rho _{p} ,\varphi _{p}} ^{} {\frac{{1}}{{m^{p}(\rho
_{p} ,\varphi _{p} )}}}\vec {\nabla} _{\rho _{p} ,\varphi _{p}}\right.\nonumber\\
&+ \left.{\frac{{1}}{{m^{p}(\rho _{p} ,\varphi _{p}
)}}}{\frac{{\partial ^{2}}}{{\partial z_{p}^{2}} }} \right] +
U^{p}\left( {\rho _{p} ,\varphi _{p}}  \right){\rm ,} \quad(p =
\mathrm{e,\,h}) 
\nonumber\\[-2ex]
\end{align}
are the Hamiltonians of non-interacting electron and hole in
cylindrical coordinates ($\rho,\varphi,z$),
\begin{equation}
\label{eq4} V(\vert \vec {r}_\mathrm{e} - \vec {r}_\mathrm{h} \vert ) = -
{\frac{{e^{2}}}{{\varepsilon (\vec {r}_\mathrm{e} ,\,\vec {r}_\mathrm{h} )\vert
\vec {r}_\mathrm{e} - \vec {r}_\mathrm{h} \vert} }}
\end{equation}
is the potential energy of their interaction.

The equation (\ref{eq1}) with Hamiltonian (\ref{eq2}) cannot be solved exactly
due to the complicated dependence of the potential energy (\ref{eq4}) and
physical characteristics ($m^{p},U^{p}$) of the quasi-particles on
geometrical parameters of hexagonal nanotube.

An approximated solution of the problem is performed within two
stages~\cite{13,14}. Considering that the energy of electron-hole
interaction is much smaller than the energy of their size
quantization,  we first obtain the latter for the model of
hexagonal nanotube and then take into account the potential of
interaction of both quasi-particles (\ref{eq4}).

Thus, the stationary Schr\"odinger equations for the non-interacting electron and
hole are solved in a cylindrical coordinate system
\begin{equation}
\label{eq5} \hat{H}^{(p)}(\vec {r}_{p} )\,\psi ^{(p)}(\vec {r}_{p}
) = E^{(p)}\,\psi ^{(p)}(\vec {r}_{p} ),\qquad (p =\mathrm{e,\,h}).
\end{equation}

 It is clear that the effective masses [$m^\mathrm{(e,h)}$] and potential
energies [$U^\mathrm{(e,h)}$] as functions of $\rho$, $\varphi$ variables
have a hexagonal symmetry in the plane perpendicular to the
nanotube axis. Thus, the variables are not separated and the
equations (\ref{eq5}) cannot be solved exactly. The approximated solution
is found within the Bethe variational method. In the Hamiltonian
(\ref{eq3}) the main term is introduced. The magnitudes of $m^\mathrm{(e,h)}$ and $U^\mathrm{(e,h)}$
in this term  are the functions of a radial variable $\rho$. I.e.,
hexagons are replaced by circles of respective radii:
$\rho_{0}$, $\rho_{1}=\rho_{0}+\Delta$,
$\rho_{2}=\rho_{0}+\Delta+h$. Within this approach, the effective
masses, potential energies of an electron and a hole as well as dielectric
constants depend on the variable in the following way
\begin{equation}
\label{eq6}  m^\mathrm{(e,\,h)} = {\left\{ {\begin{array}{l}
 {m_{\,0}^\mathrm{(e,\,h)}},  \\ \\
 {m_{\,1}^\mathrm{(e,\,h)}},
 \end{array}}\ \right.} \qquad
\quad U^\mathrm{(e,\,h)} = {\left\{ {\begin{array}{l}
 {0_{}^{}},  \\ \\
 {U_{0}^\mathrm{(e,\,h)}},
 \end{array}}\ \right.}\qquad
\quad \varepsilon = {\left\{ {\begin{array}{l}
 {\varepsilon _{0}, \quad 0 \leqslant \rho \leqslant
\rho _{0} \quad \text{and} \quad \rho _{1} \leqslant \rho \leqslant \rho
_{2}},  \\ \\
 {\varepsilon _{1}, \quad \rho _{0} \leqslant
\rho \leqslant \rho _{1} \quad \text{and} \quad \rho _{2} \leqslant \rho <
\infty}.
\end{array}}\ \right.}
\end{equation}

The differences between the respective masses ${m^\mathrm{(e,h)}(\rho,
\varphi)}$ and ${m^\mathrm{(e,h)}}(\rho)$, and potentials
${U^\mathrm{(e,h)}}(\rho, \varphi)$ and ${U^\mathrm{(e,h)}}(\rho)$, arising as a
result of approximation, are taken into account in the Hamiltonian
as a perturbation. Herein, the radius of the smallest circle
$\rho_{0}$ is considered as a variational parameter according to
Bethe method.

Now, the Hamiltonian [$\hat{H}^\mathrm{(e,h)}$] of the uncoupling electron
and hole is written as follows:
\begin{equation}
\label{eq7}
\hat{H}^{(p)}=\hat{H}_{0}^{(p)}+\Delta\hat{H}^{(p)}, \qquad (p=\mathrm{e,
h}),
\end{equation}
where
\begin{equation}
\label{eq8}
\hat{H}_{0}^{(p)}=-\frac{\hbar^{2}}{2}\left[{\vec{\nabla}^{(p)}}_{\rho,\varphi}\frac{1}{{m^{(p)}}(\rho)}
{\vec{\nabla}^{(p)}}_{\rho,\varphi}+\frac{1}{{m^{(p)}}(\rho)}\frac{\partial^{2}}{\partial
z^{2}}\right]+{U^{(p)}}(\rho)
\end{equation}
is the main part of the Hamiltonian which describes the electron
and hole with effective masses ${m^\mathrm{(e,h)}}(\rho)$ and potential
energies ${U^\mathrm{(e,h)}}(\rho)$ in a multi-shell cylindrical
nanostructure.

The correction that takes into account the difference between the exact
[$\hat{H}^{(p)}$] and approximated [$\hat{H}_{0}^{(p)}$]
Hamiltonians
\begin{eqnarray}
\label{eq9}
 \Delta\hat {H}^{(p)}&=& U^{(p)}(\rho ,\varphi ) -
U^{(p)}(\rho )
\nonumber\\
&&{}+{\frac{{\hbar ^{2}}}{{2}}}{\left\{ {\vec {\nabla} _{\rho ,\varphi}
\left[ {{\frac{{1}}{{\mu ^{(p)}(\rho )}}} - {\frac{{1}}{{\mu
^{(p)}(\rho ,\varphi )}}}} \right]\vec {\nabla} _{\rho ,\varphi}
- \left[ {{\frac{{1}}{{\mu ^{(p)}(\rho )}}} - {\frac{{1}}{{\mu
^{(p)}(\rho ,\varphi
)}}}} \right]{\frac{{\partial ^{2}}}{{\partial z_{}^{2}} }}} \right\}}
\end{eqnarray}
is further considered as a perturbation.

The Schr\"odinger equation with Hamiltonian (\ref{eq8}) is solved exactly.
Herein, we obtain analytical expressions for wave functions
[$\psi _{n_{\rho}  \,m}^{(p)} (\rho _{0} ,\vec {r}_{p} ) = \varphi
_{n_{\rho} \,m}^{(p)} \,(\rho _{0} ,\,\vec {\rho} _{p} \,)\,\exp
(\ri k_{p} \,z_{p} \,)$] and electron (hole) energy spectrum
[$E_{n_{\rho \,} m_{}} ^{(0)\,(p)} (\rho _{0} ,k)$] as a function of
$\rho_{0}$ variational parameter in zero approximation of
a perturbation method.

 Further, according to Bethe method, we
calculate the corrections of the first order to the energies of
the both quasi-particles (e, h) as functions of $\rho_{0}$
\begin{equation}
\label{eq10}
\delta E_{n_{\rho \,} m}^{(p)} \,(\rho _{0} ,k) = 6{\int\limits_{
- L / 2}^{L / 2} {\,}} \,{\sum\limits_{j = 0}^{2}
{\,\,{\int\limits_{S_{j}}  {}} } }\Psi _{n_{\rho}  m}^{(p){\rm}
\ast}  \,(\rho _{0} ,\vec {r}_{p} )\,\Delta\hat {H}\,\Psi
_{n_{\rho}  m}^{(p)} \,(\rho _{0} ,\vec {r}_{p} )\,\rd^{3}\vec
{r}_{p}\,,\qquad  (p=\mathrm{e, h}), \end{equation}
where $L$  denotes
the effective region of a quasi-particle movement along the axial
axis of a nanotube.

 We should note that due to the evident analytical
properties of a perturbation Hamiltonian~(\ref{eq9}), integration over
$\rho,\varphi$ variables in the expression~(\ref{eq10}) is performed only over
$S_{j}$ regions located between the respective hexagons and
approximating circles (shadowed regions in figure~\ref{fig1}).

Now,  without taking the
electron-hole interaction into account, exciton energy spectrum is obtained from the condition of
functional minimum
\begin{eqnarray}
\label{eq11}
\varepsilon _{n_{\rho} ^\mathrm{h} m^\mathrm{h}}^{n_{\rho} ^\mathrm{e} m^\mathrm{e}} &=& E_{g_{0}}  +
E_{n_{\rho} ^\mathrm{e} m^\mathrm{e}}^{} + E_{n_{\rho} ^\mathrm{h} m^\mathrm{h}}^{} \nonumber \\
&=& E_{g_{0}}  + \min \left[ E_{n_{\rho} ^\mathrm{e} m^\mathrm{e}}^{\left( {0} \right)}
\left( {\rho _{0}}  \right) +   \delta
E_{n_{\rho} ^\mathrm{e} m^\mathrm{e}}^{} \left( {\rho _{0}}  \right) +
E_{n_{\rho} ^\mathrm{h} m^\mathrm{h}}^{\left( {0} \right)} \left( {\rho _{0}}
\right) + \delta E_{n_{\rho} ^\mathrm{h} m^\mathrm{h}}^{}
\left( {\rho _{0}}  \right) \right] _{\rho _{0} = \bar {\rho} _{0}} \,\,,
 \end{eqnarray}
realized at $\rho_{0}=\bar{\rho_{0}}$.

The electron and hole binding energies ($\Delta E_{n_{\rho} ^\mathrm{h}
m^\mathrm{h}}^{n_{\rho} ^\mathrm{e} m^\mathrm{e}}$) in the respective states are
calculated in the following way. In the space of quantum wire
(``0'') or nanotube (``2'') there is performed an averaging of the potential
interacting energy~(\ref{eq4}) at the electron and hole wave functions
describing their movement in the plane perpendicular to the axial
axis in the system of mass center [$z=z_\mathrm{e}-z_\mathrm{h}$, \ $Z =
(z_\mathrm{e}^{} m_{0}^\mathrm{e} + z_\mathrm{h}^{} m_{0}^\mathrm{h}) \big/ ( {m_{0}^\mathrm{e} +
m_{0}^\mathrm{h}})$]
\begin{equation}
\label{eq12}
 V_{n_{\rho} ^\mathrm{h} \,m^\mathrm{h}}^{n_{\rho} ^\mathrm{e} \,m^\mathrm{e}} \left( {z}
\right) = {\frac{{e^{2}}}{{\varepsilon _{0}} }}\int {\rd\,\vec
{\rho} _\mathrm{e} \,\rd\,\vec {\rho} _\mathrm{h} {\frac{{{\left| {\varphi
_{n_{\rho} ^\mathrm{e} \,m^\mathrm{e}} \left( {\bar {\rho} _{0} ,\,\,\vec {\rho}
_\mathrm{e}}  \right)\,\varphi _{n_{\rho} ^\mathrm{h} \,m^\mathrm{h}} \left( {\bar
{\rho} _{0} ,\,\vec {\rho} _\mathrm{h}}  \right)} \right|}^{2}}}{{\sqrt
{\left( {\vec {\rho} _\mathrm{e} - \vec {\rho} _\mathrm{h}} \right)^{2} +
z^{2}}} }}}\,\,.
\end{equation}

We should note that this potential describes not only the Coulomb
interaction between the electron and hole along the Oz axis but also
``effectively'' takes it into account at the transversal plane.

Now, the Hamiltonian (\ref{eq2}) takes the form
\begin{equation}
\label{eq13}
\hat{H}_\mathrm{ex} = - {\frac{{\hbar ^{2}}}{{2M}}}{\frac{{\partial
^{2}}}{{\partial Z^{2}}}} - {\frac{{\hbar ^{2}}}{{2\mu}
}}{\frac{{\partial ^{2}}}{{\partial z^{2}}}} - V_{n_{\rho} ^\mathrm{h}
\,m^\mathrm{h}}^{n_{\rho} ^\mathrm{e} \,m^\mathrm{e}} \left( {z} \right) + \varepsilon
_{n_{\rho} ^\mathrm{h} m^\mathrm{h}}^{n_{\rho} ^\mathrm{e} m^\mathrm{e}}\,\,\,.
\end{equation}
Here,
\begin{equation}
\label{eq14} M = m_{\,0}^\mathrm{e} + m_{\,0}^\mathrm{h}\, ,  \qquad
\mu = {\frac{{m_{\,0}^\mathrm{e} \,m_{\,0}^\mathrm{h}} }{{m_{\,0}^\mathrm{e} +
m_{\,0}^\mathrm{h}} }}\end{equation}
are the effective mass of an exciton generally
moving in the longitudinal direction and its reduced
mass, respectively.

From the expression (\ref{eq13}) it is clear that the movement of the mass
centre of an exciton   along the $OZ$  axis is separated in such a
way that the energy $E_{P}$ and wave function [$\Psi _{P} (Z)$] of an exciton longitudinal movement are
as follows:
\begin{equation}
\label{eq15}E_{P} = {\frac{{P^{2}}}{{2\,M}}}\,{\rm ,}\qquad
\quad \Psi _{P} (Z) = {\frac{{1}}{{\sqrt {2\pi \,\hbar} } }}\,\exp
\left(\frac{\ri PZ}{ \hbar }\right){\rm .}
\end{equation}

The Schr\"odinger equation separately for $z$-th component cannot
be solved exactly. In order to obtain its approximated solution,
in the Hamiltonian (\ref{eq13}) we add and subtract the potential
\begin{equation}
\label{eq16} V(z) = - {\frac{{e^{2}}}{{\varepsilon _{0}}
}}\,{\frac{{1}}{{(\,\beta + \vert z\vert )}}}
\end{equation}
with variational parameter $\beta$.

Such a potential, on the one hand, has the main properties of a
potential of electron-hole interaction (\ref{eq12}) and, thus, together
with the kinetic energy of $z$-th component, provides the energy
of the bound state $E_{n_{z}}$ and, on the other hand, contrary to the potential~(\ref{eq12}) provides a rather small
magnitude within the conception of perturbation theory
\begin{equation}
\label{eq17} \Delta{V}_{n_{\rho} ^\mathrm{h} \,m^\mathrm{h}}^{n_{\rho} ^\mathrm{e}
\,m^\mathrm{e}} = {\frac{{e^{2}}}{{\varepsilon _{0}} }}\left[
{{\frac{{1}}{{\beta + \vert z\vert} }} - V_{n_{\rho} ^\mathrm{h}
\,m^\mathrm{h}}^{n_{\rho} ^\mathrm{e} \,m^\mathrm{e}} \left( {z} \right)} \right].
\end{equation}

The Schr\"odinger equation
\begin{equation}
\label{eq18} \hat{H} _{z} \,\Psi _{n_{z}}  (z) = E_{n_{z}} ^{\,}
\,\Psi _{n_{z}} (z)
\end{equation}
with the Hamiltonian
\begin{equation}
\label{eq19} \hat{H} _{z} = - {\frac{{\hbar ^{2}}}{{2\,\mu}
}}\,{\frac{{\partial ^{2}}}{{\partial z^{2}}}} -
{\frac{{e^{2}}}{{\varepsilon _{0}} }}\,{\frac{{1}}{{(\,\beta +
\vert z\vert )}}}
\end{equation}
is solved exactly~\cite{15} and the wave function is obtained as
\begin{equation}
\label{eq20} \Psi _{n_{z}}  (z) = A\,\exp \left[ - \chi \,\left(z + \beta
\right)\right]\,\,F\left[ - {\frac{{\nu }}{{2\,\chi} }};\,\,0;\,\,2\chi \left(z + \beta
\right)\right].
\end{equation}
Here,
\begin{equation}
\label{eq21} \nu = {\frac{{2\mu} }{{\hbar
^{2}}}}\,{\frac{{e^{2}}}{{\varepsilon _{0} }}}\,{\rm ,}\qquad
\quad \chi ^{2} = {\frac{{2\mu} }{{\hbar ^{2}}}}\,E_{n_{z}}
\,{\rm ,}
\end{equation}
$A$ is the normality constant and $F$  is the confluent hyper-geometrical function.

The equation (\ref{eq18}) with the Hamiltonian (\ref{eq19}) is symmetrical with
respect to the replacement $z \to -z$. Thus, its solutions should be even or odd. This brings about
two boundary conditions
\begin{equation}
\label{eq22}
\left\{
  \begin{array}{ll}
    \displaystyle\frac{\partial
\Psi _{n_{z}}  (z)}{\partial \,z}\Bigg|
_{z = 0} = 0, & \hbox{$\Psi _{n_{z}}$ -- \text{even};} \\[2ex]
 \Psi _{n_{z}}  (0) = 0, & \hbox{$\Psi
_{n_{z}}$  -- \text{odd}}
  \end{array}
\right.
\end{equation}
consistently defining the energy spectrum $E_{n_{z}}$.

Now, the exciton energy, as a function of variational parameter
$\beta$, is presented by the expression
\begin{equation}
\label{eq23} E_{n_{\rho} ^\mathrm{h} \,m^\mathrm{h}}^{n_{\rho} ^\mathrm{e} \,m^\mathrm{e}}
(n_{z},\,P ,\beta) = E_{n_{\rho} ^\mathrm{e} \,m^\mathrm{e}}^{\,} + E_{n_{\rho}
^\mathrm{h} \,m^\mathrm{h}}^{\,} + {\frac{{P^{2}}}{{2\,M}}} + E_{g\,0} + \Delta
E_{n_{\rho} ^\mathrm{h} \,m^\mathrm{h}}^{\,n_{\rho} ^\mathrm{e} \,m^\mathrm{e}} (n_{z} ,\beta
)
\end{equation}
and the wave functions of zero approximation
\begin{equation}
\label{eq24} \Psi _{n_{\rho} ^\mathrm{h} \,m^\mathrm{h}}^{n_{\rho} ^\mathrm{e} \,m^\mathrm{e}}
(n_{z} ,\,P,\,\beta ) = \Psi _{P} (Z)\,\Psi _{n_{z}}  (\beta
,\,z)\,\varphi _{n_{\rho} ^\mathrm{e} \,m^\mathrm{e}}^{} \,(\bar{\rho
_{0}},\,\rho_\mathrm{e} ,\,\varphi _\mathrm{e} )\,\varphi _{n_{\rho} ^\mathrm{h}
\,m^\mathrm{h}}^{} \,(\bar{\rho _{0}},\,\rho_\mathrm{h} ,\,\varphi _\mathrm{h} ) =
{\left| {{\begin{array}{*{20}c}
 {n_{\rho} ^\mathrm{e} \,m^\mathrm{e}} \hfill \\
 {n_{\rho} ^\mathrm{h} \,m^\mathrm{h}} \hfill \\
\end{array}} n_{z} \,P} \right\rangle}.
\end{equation}
The binding energy [$\Delta E_{n_{\rho} ^\mathrm{h} \,m^\mathrm{h}}^{\,n_{\rho}
^\mathrm{e} \,m^\mathrm{e}} (n_{z} ,\beta )$] of an exciton in the expression~(\ref{eq23}),
naturally, consists of the energy of the bound state $E_{n_{z}}$
along $Oz$ axis and the correction [$\delta E_{n_{\rho} ^\mathrm{h}
\,m^\mathrm{h}}^{\,n_{\rho} ^\mathrm{e} \,m^\mathrm{e}} (n_{z} ,\beta)$] calculated
as a diagonal matrix element of a perturbation operator (\ref{eq17}) at the
wave functions (\ref{eq24})
\begin{eqnarray}
\label{eq25} \Delta E_{n_{\rho} ^\mathrm{h} \,m^\mathrm{h}}^{\,n_{\rho} ^\mathrm{e}
\,m^\mathrm{e}} (n_{z} ,\beta ) & = & E_{n_{z}}  (\beta ) + \delta
E_{n_{\rho} ^\mathrm{h} \,m^\mathrm{h}}^{\,n_{\rho} ^\mathrm{e} \,m^\mathrm{e}} (n_{z} ,\beta
),\\
%
\label{eq26} \delta E_{n_{\rho} ^\mathrm{h} \,m^\mathrm{h}}^{\,n_{\rho} ^\mathrm{e}
\,m^\mathrm{e}} (n_{z} ,\beta ) & = & {\frac{{e^{2}}}{{\varepsilon _{0}}
}}{\left\langle {n_{z} \,} \right|}\left( {{\frac{{1}}{{\beta +
\vert z\vert} }} - V_{n_{\rho} ^\mathrm{h} \,m^\mathrm{h}}^{n_{\rho }^\mathrm{e}
\,m^\mathrm{e}} \left( {z} \right)} \right){\left| {n_{z} \,}
\right\rangle }\,\,\,.
\end{eqnarray}

Having the magnitude $\bar {\beta}$ ensuring the minimum of
$\Delta E_{n_{\rho} ^\mathrm{h} \,m^\mathrm{h}}^{\,n_{\rho} ^\mathrm{e} \,m^\mathrm{e}} (n_{z}
,\bar {\beta} ) = \Delta E_{n_{\rho} ^\mathrm{h} \,m^\mathrm{h}}^{\,n_{\rho}
^\mathrm{e} \,m^\mathrm{e}} (n_{z} )$, we obtain final expressions for the
energies
\begin{equation}
\label{eq27} E_{n_{\rho} ^\mathrm{h} \,m^\mathrm{h}}^{n_{\rho} ^\mathrm{e} \,m^\mathrm{e}}
(n_{z} ,\,P) = E_{n_{\rho }^\mathrm{e} \,m^\mathrm{e}}^{\,} + E_{n_{\rho} ^\mathrm{h}
\,m^\mathrm{h}}^{\,} + {\frac{{P^{2}}}{{2\,M}}} + E_{g\,0} + \Delta
E_{n_{\rho} ^\mathrm{h} \,m^\mathrm{h}}^{\,n_{\rho} ^\mathrm{e} \,m^\mathrm{e}} (n_{z} )
\end{equation}
and wave functions
\begin{equation}
\label{eq28} \Psi _{n_{\rho} ^\mathrm{h} \,m^\mathrm{h}}^{n_{\rho} ^\mathrm{e} \,m^\mathrm{e}}
(n_{z} ,\,P) = \Psi _{P} (Z)\,\Psi _{n_{z}}  (z)\,\varphi
_{n_{\rho} ^\mathrm{e} \,m^\mathrm{e}}^{} \,(\bar{\rho_{0}},\,\rho _\mathrm{e}
,\,\varphi _\mathrm{e} )\,\varphi _{n_{\rho} ^\mathrm{h} \,m^\mathrm{h}}^{}
\,(\bar{\rho_{0}},\,\rho _\mathrm{h} ,\,\varphi _\mathrm{h} )
\end{equation}
of the exciton in a multi-shell hexagonal nanotube.

The electron and hole wave functions are used for the evaluation
of the intensities of the interband optical quantum transitions
according to the formula~\cite{16}
\begin{equation}
\label{eq29} I_{n_{\rho} ^\mathrm{h} m^\mathrm{h}}^{n_{\rho} ^\mathrm{e} m^\mathrm{e}} \sim
{\left| {\int\!\!\!\int {\varphi _{n_{\rho} ^\mathrm{e} \,m^\mathrm{e}}^{\ast}
\,(\bar{\rho_{0}},\,\vec {\rho} )\varphi _{n_{\rho} ^\mathrm{h} \,m^\mathrm{h}}
\,(\bar{\rho_{0}},\,\vec {\rho})\,\,\rd\vec {\rho} } \,}
\right|}^{2}.
\end{equation}

A further calculation of the spectrum and the analysis of its
properties was performed using numeric methods for the     GaAs/Al$_{0.4}$Ga$_{0.6}$As multi-shell hexagonal nanotube grown experimentally~\cite{8}.

\section{Discussion of results}

Exciton spectrum as a function of a nanostructure geometrical
parameters is studied for \linebreak GaAs/Al$_{0.4}$Ga$_{0.6}$As
multi-shell hexagonal nanotube having physical parameters~\cite{7,8,17}:
$m_{0}^\mathrm{e}=0.063 \, m_{0}$,  $m_{1}^\mathrm{e}=0.096 \,
m_{0}$, $m_{0}^\mathrm{h}=0.51 \, m_{0}$, $m_{1}^\mathrm{h}=0.61 \,m_{0}$,
$U_{0}^\mathrm{e}=297$~meV, $U_{0}^\mathrm{h}=224$~meV, $E_{g\,0}=1520$~meV,
$\varepsilon_{0}=10.89$, ($m_{0}$ is the pure electron mass in
vacuum); $a_\mathrm{GaAs}=5.65$~\AA \ is the lattice
constant of GaAs.

\begin{figure}[!h]
\centerline{
\includegraphics[width=0.68\textwidth]{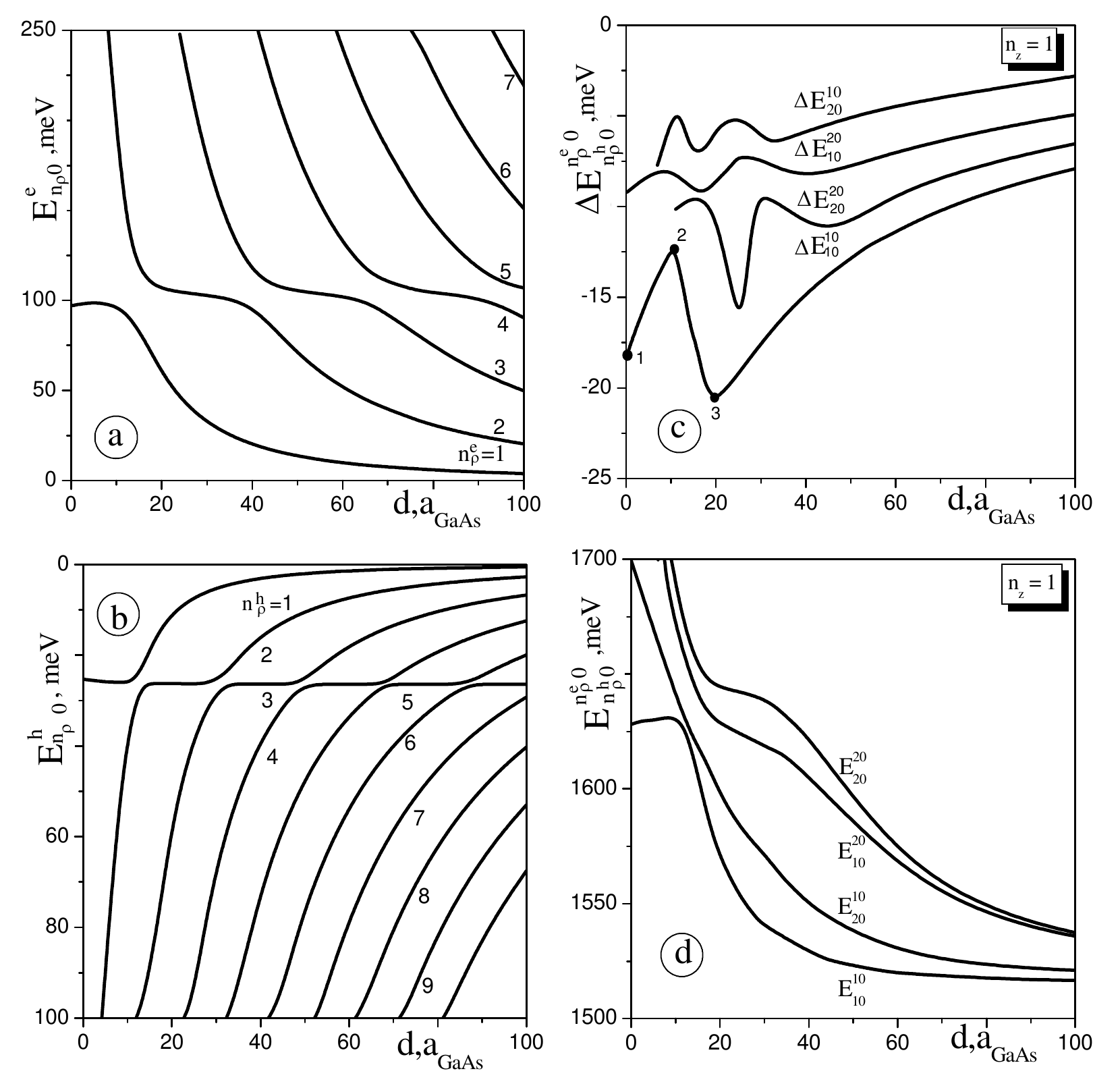}
}
\caption{Dependences of electron energy $E_{n_{\rho} 0}^\mathrm{e}$ (a),
heavy hole energy $E_{n_{\rho} 0}^\mathrm{h}$ (b), exciton binding energy
$\Delta E_{n_{\rho} ^\mathrm{h} \,0}^{\,n_{\rho} ^\mathrm{e} \,0}$ (c) and
exciton energy $E_{n_{\rho} ^\mathrm{h} \,0}^{\,n_{\rho} ^\mathrm{e} \,0}$(d) on
the inner wire diameter (d) at $P=0$, $ n_{z}=1$, $m=0$, and
the experimentally obtained barrier-shell width ($\Delta
=4$~nm) and nanotube width ($h=4$~nm)~\cite{5}. Points 1, 2, 3 in figure~\ref{fig2}~(c)
denote the extreme magnitudes of exciton binding energy:
point 1~-- electron and hole are located in the space of nanotube;
point 2~-- the hole is located in nanotube and the electron~-- in
nanotube and the inner wire with the probabilities $\approx 0,5$;
point 3~-- the both electron and hole, with probability close to
one, are located in the inner wire.}\label{fig2}
\end{figure}

In figure~\ref{fig2}, the electron energy $E_{n_{\rho}  0}^\mathrm{e}$ (a), heavy
hole energy $E_{n_{\rho}  0}^\mathrm{h}$ (b), exciton binding energy
$\Delta E_{n_{\rho} ^\mathrm{h} \,0}^{\,n_{\rho} ^\mathrm{e} \,0}$ (c) and
exciton energy $E_{n_{\rho} ^\mathrm{h} \,0}^{n_{\rho} ^\mathrm{e} \,0}$ (d) are
presented as functions of the inner wire GaAs diameter $d$ at $P =
0$, $n_{z} = 1$, and experimental magnitudes for the
barrier-shell size: $\Delta=4$~nm and nanotube width: $h=4$~nm~\cite{8}. In figures~\ref{fig2}~(a), (b) one can see only two energy levels at
$d=0$. These levels coincide with the ones obtained for a hexagonal
nanotube (GaAs) embedded into Al$_{0.4}$Ga$_{0.6}$As, which is
proven by physical considerations.

The new electron and hole energy levels arise when the quantum
wire appears and its diameter $d$ increases. The whole spectra
shift into the region of lower energies and the anti-crossings of
energy levels are observed. The anti-crossing phenomena are caused
by the splitting of energy levels due to the tunnel effect present
between the quantum wire (with diameter $d$) and the nanotube (of $h$
width) through the finite potential barrier (of $\Delta$ width). Both the electron and hole are located in the space of a nanotube at that plots of  $E_{n_{\rho}  0}^\mathrm{(e,\,h)}$ dependences
on $d$, where the energies of quasi-particles are almost unchanged. The plots where the energies of both quasi-particles rapidly decrease correspond to the states in which the electron
and hole are located in the inner wire with the probability close to
one.

The exciton binding energies ($\Delta E_{n_{\rho} ^\mathrm{h}
\,0}^{\,n_{\rho} ^\mathrm{e} \,0}$) [figure 2 (c)] non-monotonously
depend on the inner wire diameter $d$ taking several minimal and
maximal magnitudes for all states. This is clear because when the
electron and hole are in their ground states, then, at $d=0$, there is no
inner wire, and the both quasi-particles are localized
in the space of a nanotube (with $h=4$~nm) and  the binding energy is
$\Delta E_{\,1\,0}^{\,1\,0} \approx18$~meV [point ``1'' in figure~\ref{fig2}~(c)].

When the inner wire appears and its diameter increases, the
absolute value of the binding energy somewhat decreases
because the electron wave function, as the one for the light
quasi-particle, increasingly penetrates into the space of the inner
wire, while a massive hole does not change its location. Herein,
the effective distance between quasi-particles increases. At some
critical $d$ [point ``2'' in figure~\ref{fig2}~(c)], the massive hole also
begins to penetrate into the inner wire. The effective distance
between quasi-particles decreases and, consequently, the absolute
value of the binding energy increases. The maximal value of the
binding energy ($\Delta E_{\,1\,0}^{\,1\,0} \approx 22$~meV) is
obtained when  both the electron and hole, with probability close
to one, are located in the inner wire [point ``3'' in figure~\ref{fig2}~(c)]. At
a further  increase of $d$, the binding energy decreases only, because
the effective distance between the electron and hole in the space of
the inner wire becomes bigger.
\begin{figure}[!h]
\centerline{
\includegraphics[width=0.68\textwidth]{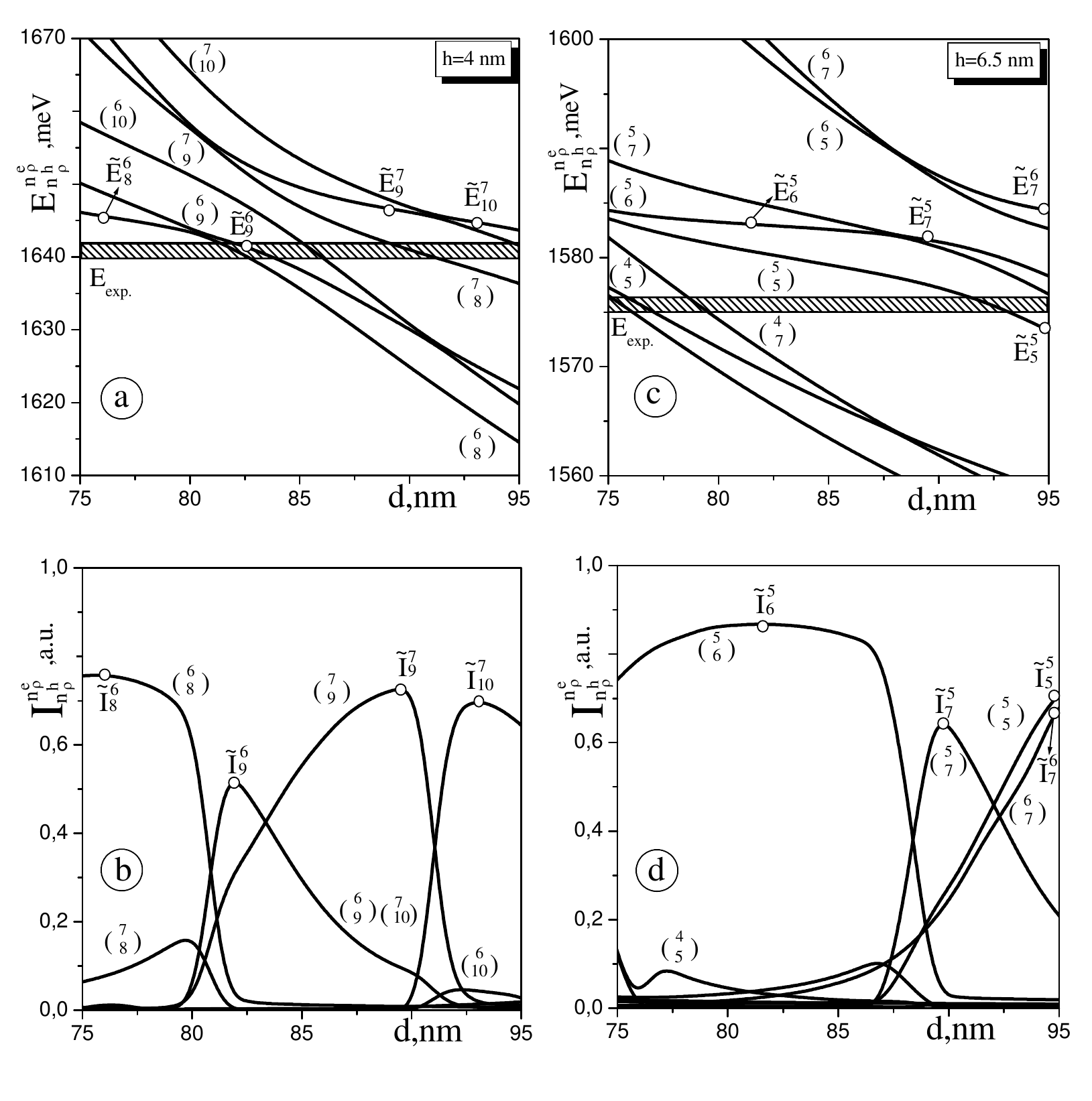}
}
\caption{Dependences of exciton energies $E_{n_{\rho} ^\mathrm{h}}
^{\,n_{\rho} ^\mathrm{e}}$ and intensities of inter-band quantum
transitions $I_{n_{\rho} ^\mathrm{h}} ^{\,n_{\rho} ^\mathrm{e}}$ on inner wire
diameter (d) in the range: $d=85\pm10$~nm at nanotube widths:
$h_{1}=4$~nm (a, b) and $h_{2}=6.5$~nm (c, d). The magnetic
quantum number is equal to zero, thus index ``0'' is omitted. The
region of energies where the radiation peaks are experimentally
observed~\cite{8} is shown at the figures by dash strips. The maximal
magnitudes of the intensities of quantum transitions and the
respective exciton energies are shown by the points.}\label{fig3}
\end{figure}

Similarly, the non-monotonous behaviour of a binding energy of an exciton in its
excited states can be explained by the change of an
electron and hole location in a quantum wire or nanotube.

\newpage
We should note that $V(z)$ potential (\ref{eq16}) insomuch well
approximates the effective potential (\ref{eq12}) that the corrections to
the exciton binding energy (\ref{eq26}) do not exceed 0.5~meV in all
states and for any geometrical sizes of a nanotube.

The exciton energy is two orders bigger than the absolute value of
the binding energy. Thus, the dependences of the exciton energies
$E_{n_{\rho} ^\mathrm{h} \,0}^{\,n_{\rho} ^\mathrm{e} \,0}$ in the region of low
energies on the inner wire diameter $d$, figure~\ref{fig2}~(d), are completely
determined by the peculiarities of electron and hole energies. In
particular,  the anti-crossing of exciton energies
is observed in these functions. This is caused by anti-crossings of electron and
hole energy levels.

In order to compare the obtained theoretical results with the
experimental ones, in figure 3, the exciton energies $E_{n_{\rho}
^\mathrm{h}} ^{\,n_{\rho} ^\mathrm{e}}$ and respective intensities of quantum
transitions $I_{n_{\rho} ^\mathrm{h}} ^{\,n_{\rho} ^\mathrm{e}}$ are presented
as functions of the inner wire diameter $d$ within the ranges: $d =85\pm10$~nm and for two nanotube widths:
$h_{1}=4$~nm [figure\ref{fig3}~(a), (b)] and $h_{2}=6.5$~nm [figure~\ref{fig3}~(c),
(d)]. At these parameters, the luminescence peaks with the
energies shown in figures~\ref{fig3}~(a), (c) by dashed strips were
observed experimentally. The numerical calculations of exciton
energies and intensities of quantum transitions are performed at a
magnetic quantum number equal to zero, thus index ``0'' is dropped
for convenience.

All figures prove that only some of the exciton curves in the
vicinity of experimental data can be characterized by the
intensities more than 0.5. For example at  $h_{1}=4$~nm these
energies are: $ E_{8}^{6}$, $E_{9}^{6}$, $E_{9}^{7}$, $E_{10}^{7}$ [figure~\ref{fig3}~(a), (b)] and
$E_{5}^{5}$, $E_{6}^{5}$, $E_{7}^{5}$, $E_{7}^{6}$ at $h_{2}=6.5$~nm [figure\ref{fig3}~(c), (d)]. The
maximal intensities ($\tilde {I}_{n_{\rho} ^\mathrm{h}} ^{n_{\rho}
^\mathrm{e}}$) correspond to the exciton energies ($\tilde {E}_{n_{\rho}
^\mathrm{h}} ^{n_{\rho} ^\mathrm{e}}$), at which, evidently, the luminescence
peaks should be experimentally observed. Then, depending on the
exact experimental sizes of the inner wire, the luminescence peaks
observed experimentally~\cite{8} can be produced by different exciton
states with the energies slightly different from each other at a
fixed nanotube width $h$.

Finally, the calculations of probability density of electron and
hole location in a nanostructure prove that the both
quasi-particles producing the exciton in the above mentioned states
($\tilde {E}_{n_{\rho} ^\mathrm{h}} ^{n_{\rho} ^\mathrm{e}}$) are localized in
the nanotube of  $h$ width.

\section{Conclusions}

\begin{enumerate}

\item The theory of exciton spectra for a multi-shell hexagonal
nanotube is developed within the models of effective masses and
rectangular potentials for the electron and hole using the method
of the effective potential for obtaining the exciton binding
energy.

\item The exciton binding energy for all states non-monotonously
depends on the inner wire diameter $d$ approaching several minimal
and maximal magnitudes. The behaviour of the exciton binding energy is
quite well explained by the complicated character of distributions of
probability densities of the electron and hole locations in the
spaces composition parts of a multi-shell nanotube.

\item The numerical results for the exciton energy spectrum and
intensities of quantum transitions, obtained within the developed
theory, correlate well with the experimental data for the radiation
spectra of GaAs/Al$_{0.4}$Ga$_{0.6}$As nanotubes.

\end{enumerate}



\newpage

\begin{thebibliography}{17}

\bibitem{1} Suenaga K., Colliex C., Demoncy N., Loiseau A., Pascard H., Willaime F., Science, 1997, \textbf{278}, 653; \doi{10.1126/science.278.5338.653}.

\bibitem{2} Zhang Y., Suenaga K., Colliex C., Iijima S., Science, 1998, \textbf{281}, 973; \doi{10.1126/science.281.5379.973}.

\bibitem{3} Persson A.I., Larsson M.W., Stenstro S., Ohlsson B.J., Samuelson L., Wallenberg L.R.,
 Nat. Mater., 2004, \textbf{3}, 677; \doi{10.1038/nmat1220}.

\bibitem{4} Dubrovskii V.G., Cirlin G.E.,  Ustinov V.M., Semiconductors, 2009, \textbf{43}, No~12, 1539; \doi{10.1134/S106378260912001X}.

\bibitem{5} Mohan P., Motohisa J., Fukui T., Appl. Phys. Lett., 2006, \textbf{88}, 013110; \doi{10.1063/1.2161576}.

\bibitem{6} Mohan P., Motohisa J., Fukui T., Appl. Phys. Lett., 2006, \textbf{88}, 133105; \doi{10.1063/1.2189203}.

\bibitem{7} Heigoldt M., Arbiol J., Spirkoska D.,  Rebled J.M., Conesa-Boj~C.S., Abstreiter~G., Peiro~F.,  Morantece~J.R., Fontcuberta i Morral~A., J. Mater. Chem., 2009, \textbf{19}, 840; \doi{10.1039/b816585h}.

\bibitem{8} Fontcuberta i Morral A., Spirkoska D.,  Arbiol J., Heigoldt  M., Morante J.R., Abstreiter G.,  Small, 2008, \textbf{4}, 899; \doi{10.1002/smll.200701091}.

\bibitem{9} Kasapoglu E., Sari  H., Sokmen I., Surf. Rev. Lett., 2003, \textbf{10},  737; \doi{10.1142/S0218625X03005566}.

\bibitem{10} Bouhassoune  M., Charrour R., Fliyou M., J. Appl. Phys., 2002, \textbf{91}, 232; \doi{10.1063/1.1419261}.

\bibitem{11} Sidor Y., Partoens B., Peeters  F.M., Phys. Rev. B, 2007, \textbf{76}, 195320; \doi{10.1103/PhysRevB.76.195320}.

\bibitem{12} Slachmuylders A.F.,  Partoens B.,  Magnus W., Physica E, 2008, \textbf{40}, 2166; \doi{10.1016/j.physe.2007.10.091}.

\bibitem{13} Tkach M., Makhanets O., Dovganiuk M., Phys. Solid State, 2009, \textbf{51}, 2529; \doi{10.1134/S1063783409120166}.

\bibitem{14} Tkach M., Makhanets O., Dovganiuk  M., Voitsekhivska  O., Physica E, 2009, \textbf{41}, 1469; \\ \doi{10.1016/j.physe.2009.04.018}.

\bibitem{15} Ogawa  T., Takagahara T., Phys. Rev. B, 1991, \textbf{44}, 8138; \doi{10.1103/PhysRevB.44.8138}.

\bibitem{16} Davies J.H., The Physics of Low-dimensional Semiconductors: An
Introduction. Cambridge University Press, New York, 1998.

\bibitem{17} Hai G.Q., Peeters F.M., Devreese  J.T., Phys. Rev. B, 1993, \textbf{48}, 4666; \doi{10.1103/PhysRevB.48.4666}.

\end{thebibliography}

\ukrainianpart

\title{Екситонний спектр у багатошаровій шестигранній напівпровідниковій нанотрубці}

\author{O.M. Маханець, В.І. Гуцул, Н.Р. Цюпак, O.M. Войцехівська}
\address{Чернівецький національний університет ім.Ю.~Федьковича, \\Україна, 58012 Чернівці, вул.Коцюбинського, 2}

\makeukrtitle

\begin{abstract}
\tolerance=3000%
У наближенні ефективних мас та прямокутних потенціалів, з
використанням методу ефективного потенціалу побудовано теорію
екситонного спектра у складній багатошаровій шестигранній
напівпровідниковій нанотрубці. Отримані теоретичні результати
добре пояснюють експериментальні положення піків люмінесценції у
нанотрубках GaAs/Al$_{0.4}$Ga$_{0.6}$As.

\keywords шестигранна
нанотрубка, квантовий дріт, екситонний спектр

\end{abstract}

\end{document}